\documentclass[11pt,a4paper]{article}

\usepackage{jcappub}
\pdfoutput=1 

\usepackage[T1]{fontenc}
\usepackage{subcaption}
\usepackage{amsmath}
\usepackage{multirow}
\usepackage{amssymb}
\usepackage{mathrsfs} 
\usepackage{time}
\usepackage{relsize, comment}
\usepackage[utf8]{inputenc}

\title{Probing  the warped vacuum geometry around a Kerr black hole by quasi-periodic oscillations}

\author[a]{M. Hossein Hesamolhokama}
\emailAdd{hesamolhokama@khu.ac.ir}

\author[a]{Alireza Allahyari 
}
\emailAdd{alireza.al@khu.ac.ir}

\author[b]{Jafar Khodagholizadeh}
\emailAdd{gholizadeh@ipm.ir}

\author[a]{Ali Vahedi }
\emailAdd{vahedi@khu.ac.ir}

\affiliation[a]{Department of Astronomy and High Energy Physics, Kharazmi University, 15719-14911, Tehran, Iran \looseness=-1}
\affiliation[b]{Department of Physics Education, Farhangian University, P.O. Box 14665-889, Tehran, Iran}

\abstract{We investigate quasi-periodic oscillations (QPOs) in the context of a new rotating black hole solution that incorporates a cosmological constant. Recent work by the authors in \cite{Ovalle:2022eqb} interpreted the cosmological constant, denoted as $\Lambda$, as a form of vacuum energy and employed a gravitational decoupling approach to derive an extended Kerr–de Sitter black hole solution, which is geometrically richer than the classical case. In this study, we derive the expressions for timelike circular geodesics within this solution and, using a relativistic precision model, calculate the corresponding frequencies of the QPOs.

To constrain our model, we apply Bayesian formalism, utilizing data from three well-known microquasars: GRO 1655-40, XTE 1550-564, and GRS 1915+105. Our analysis reveals that$\Lambda$
 is degenerate and correlated with other parameters. Finally, we perform a Bayesian model comparison with the Kerr metric and find that the Kerr metric is favored among the models considered.}

\begin{document}
\maketitle
\flushbottom
\section{Introduction}
Studying black holes in an expanding universe is one of the most significant challenges in cosmology \cite{Condon:2018eqx}. Black holes reside at the centers of galaxies and can accelerate matter in X-ray binaries, leading to X-ray emissions. Notably, the direct observation of the black hole shadow in $M87^\star$
 by the Event Horizon Telescope (EHT) \cite{EventHorizonTelescope:2019dse} highlights the importance of this topic and provides a valuable opportunity to test various physical theories.

The Kerr–de Sitter metric is commonly used to describe rotating black holes in an expanding universe. This solution was first derived by Carter \cite{kerrds2} and belongs to a broader class of metrics known as Kerr–Schild metrics \cite{Kerr:1965vyg}. In this class, the Einstein equations take the form of linear differential equations, allowing for the straightforward generation of new solutions.

In this context, a metric distinct from Carter's has been obtained using the method of gravitational absorption, referred to as the extended Kerr–de Sitter metric \cite{Ovalle:2021jzf}. This new metric yields a variable Ricci scalar curvature that asymptotically approaches the constant value of the Kerr–de Sitter metric at infinity. Consequently, it qualifies as an asymptotically de Sitter solution.

The specific energy, angular momentum, and Keplerian angular velocity of particles on equatorial circular orbits in the extended Kerr–de Sitter spacetime are analyzed for both circular and epicyclic motion of test particles along these orbits \cite{Slany:2023ndt}. There are quantitative limits on the existence and stability of circular orbits based on the values of energy and angular momentum for a given orbit, which could, in principle, be tested. The main differences between the two spacetimes, standard Kerr–de Sitter (KdS) and extended Kerr–de Sitter, lie in the location of the inner marginally stable orbit (determined by the locations of local minima in the profiles) and in the accretion efficiency, denoted as $\eta$
, which is given by the difference in specific energies of particles on the outer orbit (determined by the location of the local maximum) and the inner marginally stable orbit. Additionally, in this spacetime, the accretion efficiency for corotating orbits is larger than that in the standard KdS spacetime.By employing Jacobi elliptic functions, analytical solutions have been derived for both bound and nearly bound photon orbits in KdS and extended KdS spacetimes\cite{Omwoyo:2022qbk}. For KdS, large values of $ \Lambda $ increase the rate of exponential deviation from the equatorial circular prograde photon orbits, while decreasing $\Lambda$ will decrease the rate of exponential deviation from the equatorial circular retrograde photon orbits. On the other hand, for extended Kerr-de Sitter metric, increasing $  \Lambda$ increases the rate of exponential deviation from both the
equatorial circular prograde and retrograde photon orbits.\\
It is shown in \cite{Ovalle:2021jzf}  that there is a systematic way to explore the effect  a polytropic fluid
has on an arbitrary source through the
gravitational decoupling \cite{Ovalle:2017fgl} and the effect of a polytrope on the Tolman $VII$
geometry is studied~\cite{Contreras:2022nji}.  The interesting result is that $  \Lambda$
could acquire local properties in the presence of matter, and non-uniform vacuum energy does not even require the existence of any
additional form of matter-energy  \cite{Ovalle:2022eqb}. Moreover, the shadow formula of the extended Kerr-de Sitter has been derived and then visualized for various parameter ranges for an observer at given coordinates  \cite{Afrin:2021ggx}. Generally,
 an increase in the cosmological constant decreases the size of a KdS and an extended Kds black hole
shadow \cite{Omwoyo:2021uah}. 

 An arena to test gravity near black holes is by quasi-periodic oscillations observed in the radiation from the X-ray binaries \cite{Ingram:2019mna,Abramowicz:2001bi,Pasham:2014ybe,Pasham:2018bkt,Belloni_2012}. It provides an environment to test the gravity in strong field regimes \cite{Berti:2015itd,Psaltis:2008bb,Psaltis_2009,Jusufi:2020odz,Azreg-Ainou:2020bfl} and obtains parameters of the black hole and test the No-Hair Theorem \cite{Smith:2020npb,Aschenbach:2004kj,Tursunov:2018wgx,Kolos:2017ojf,Stuchlik:2013esa,Abramowicz:2004je,Johannsen:2010bi,Johannsen:2016uoh,Johannsen:2012ng,Allahyari:2021bsq}. QPOs can be measured with high precision; therefore, they are valuable for studying compact objects
\cite{Maselli:2017kic,Franchini:2016yvq,Suvorov:2015yfv,Boshkayev:2015mna,Maselli:2014fca,Cardoso:2019rvt}. There are various explanations for this phenomenon but one motivated model is the relativistic precession model in which the QPOs arise from the perturbations of matter near the last circular orbits in the vicinity of the black holes. \\
Our aim is to test the extended Kds solutions using QPOs data. We first derive the geodesics for matter within this geometry and determine the frequency of deviations from circular orbits. This allows us to assess the effect of the cosmological constant on the geodesics. We employ a Bayesian approach to fit our model to the QPOs data, leading us to establish constraints on the model's parameters. \\

 This paper is organized as follows. Section~\ref{II} presents the new Kerr-de Sitter geometry. In Section~\ref{III}, we derive the expressions for the QPOs within this geometry and discuss the relativistic precession model, including plots of the corresponding frequencies. Section~\ref{qpo} utilizes various data sets to constrain the models. Finally, we compare the models in Section~\ref{V}.

\section{Extended Kerr-de Sitter metric}\label{II}
	There exist two models for the description of the Kerr-de Sitter universe. 
For the first time, the Kerr-de Sitter metric was discovered by Carter in 1973 \cite{Carter:1970ea}. The metric in the Boyer-Lindquist coordinates is given by
	\begin{equation}
		\begin{split}
			ds^2 &= \left( \frac{\Delta_r - \Delta_\theta a^2 \sin^2(\theta)}{\rho^2\  \Xi^2} \right) dt^2 - \frac{\rho^2}{\Delta_r} dr^2 -\frac{\rho^2}{\Delta_\theta} d\theta^2  \\
			&- \frac{\sin^2(\theta)}{\rho^2\ \Xi^2}\left( \Delta_{\theta} (r^2 + a^2)^2 - \Delta_r  a^2 \sin^2(\theta)\right) d\phi^2 + \frac{2 a \sin^2(\theta) }{\rho^2\ \Xi^2}\left( \Delta_{\theta} (r^2 + a^2) - \Delta_r \right) dt d\phi,
		\end{split}
	\end{equation}
	where we have
	\begin{equation}
	\begin{split}
		\Delta_r &= r^2 - 2 M r + a^2 - \frac{\Lambda}{3} r^2 (r^2  + a^2),\\
		\Delta_\theta &= 1+ \frac{\Lambda}{3} a^2 \cos^2(\theta),\\
		\Xi &= 1+ \frac{\Lambda}{3} a^2.
	\end{split}
	\end{equation}
	This solution is a special case of the general Plebanski-Demianski family of metrics \cite{Plebanski:1976gy}. 
 
 Recently, by interpreting the cosmological constant as the energy of the vacuum, a new metric has been derived, described by the following relation:
	\cite{Ovalle:2021jzf}
	\begin{equation}\label{eq:new kerr-ds metric}
		ds^2=\left( \frac{\Delta_\Lambda-a^2 sin^2\theta}{\rho^2}\right) dt^2-\frac{\rho^2}{\Delta_\Lambda}dr^2-\rho^2 d\theta^2-\frac{\Sigma_\Lambda sin^2\theta}{\rho^2}d\phi^2+\frac{2a sin^2\theta}{\rho^2}(r^2+a^2-\Delta_\Lambda)dt d\phi,
	\end{equation}
	with 
	\begin{eqnarray}
	\Delta_\Lambda &=&r^2-2Mr+a^2-\frac{\Lambda}{3}r^4 ,\nonumber\\
	\Sigma_\Lambda &=&(r^2+a^2)^2-\Delta_\Lambda a^2 sin^2\theta, \\
	 \rho^2 &=&r^2+a^2 cos^2\theta. \nonumber 
	\end{eqnarray}
 Here $M$ and $a=\dfrac{J}{M}$ are the mass and the angular momentum density of the central rotating body respectively. This is the new solution of the Einstein field equation describing the exterior of a rotating black hole object in de Sitter or anti-de Sitter background.\\
\section{Frequency of quasi-periodic oscillation }\label{III}
According to the relativistic precession model, the quasi-periodic oscillations (QPOs) observed in X-ray binaries occur when matter orbiting the black hole in circular paths experiences small perturbations. These perturbations have characteristic frequencies that are related to the QPO frequencies.

Let's consider a test particle orbiting around a stationary, axisymmetric source. The overall line element in spherical coordinates is expressed as follows:\cite{Vahedi:2021ssf,Abramowicz:2004tm,Kerner:2001cw}
\begin{equation}
ds^2 = g_{tt} dt^2 + g_{rr}dr^2 + g_{\theta\theta} d\theta^2 
+ 2g_{t\phi}dt d\phi + g_{\phi\phi}d\phi^2 \ ,
\end{equation}
All metric components depend on $r$ and $\theta$. Thus, there are two killing vectors along the time $t$ and $\phi$, in other words, we have two constants of motion imposed by the symmetries of the metric, the specific energy at infinity, $E$ and the $z$ component of the specific angular momentum $L_z$. Therefore, we have 
\begin{eqnarray}
\dot{t} &= \frac{E g_{\phi\phi} + L_z g_{t\phi}}{
	g_{t\phi}^2 - g_{tt} g_{\phi\phi}} \,\label{eqt} , \quad \\
\dot{\phi}& = - \frac{E g_{t\phi} + L_z g_{tt}}{
	g_{t\phi}^2 - g_{tt} g_{\phi\phi}} \,\label{eqfi} \; ,
\end{eqnarray}
here dot represents the derivative with respect to the affine parameter.
For a better comparison of extended KdS (eKds) metric and Carter Kerr-dS metric  geodesic equations \cite{Vahedi:2021ssf}, we present the  expressions   for both  as follows
\begin{subequations}
	\begin{equation}
		\dot{t}_{eKdS}=\frac{\left[(r^2+a^2)^2-a^2\Delta_\Lambda\right]E+\left[a\Delta_\Lambda-(r^2+a^2)a\right]L_z}{\mu r^2 \Delta_\Lambda},
	\end{equation}
	\begin{equation}	
		{\frac{\dot{t}_{Carter}}{\chi^2}=\frac{\left[(r^2+a^2)^2-a^2\Delta_r\right]E+\left[a\Delta_r-(r^2+a^2)a\right]L_z}{\mu r^2 \Delta_r}},
	\end{equation}
 \end{subequations}

	\begin{subequations}
	\begin{equation}
		\dot{\phi}_{eKdS}=\frac{\left[ar^2+a^3-\Delta_\Lambda a\right]E+\left[\Delta_\Lambda-a^2\right]L_z}{\mu\Delta_\Lambda r^2},
	\end{equation}
	\begin{equation}
		{\frac{\dot{\phi}_{Carter}}{\chi^2}=\frac{\left[ar^2+a^3-\Delta_r a\right]E+\left[\Delta_r-a^2\right]L_z}{\mu\Delta_r r^2}},
	\end{equation} 	
 \end{subequations}
\begin{subequations}
	\begin{equation}
		\dot{r}_{eKdS}=\sqrt{\frac{\left[(r^2+a^2)^2-a^2\Delta_\Lambda\right]E^2+\left[a^2-\Delta_\Lambda\right]L_z^2+\left[2\Delta_\Lambda-2(r^2+a^2)\right]aEL_z-\mu^2r^2\Delta_\Lambda}{\mu^2 r^4}},
	\end{equation}
	\begin{equation}
	\dot{r}_{Carter}=\sqrt{\frac{\left[(r^2+a^2)^2-a^2\Delta_r\right]E^2+\left[a^2-\Delta_r\right]L_z^2+\left[2\Delta_r-2(r^2+a^2)\right]aEL_z-\mu^2 r^2\Delta_r}{\mu^2 r^4}},
	\end{equation}	
\end{subequations}
 where
	\begin{equation}
		\begin{split}
		&\Delta_r=(r^2+a^2)(1-\frac{\Lambda r^2}{3})-2Mr, \\
		&\chi^2=1+\frac{\Lambda a^2}{3},
		\end{split}
	\end{equation}
 and 
 $\mu$ is the mass of the test particle in these metrics. The most obvious difference is that Carter's metric geodesic equations have terms of 
	$\Lambda^2\,r^2$ while the new metric geodesic equations do not have these terms.
For a time-like path ; $g_{\mu\nu}\dot{x}^\mu \dot{x}^\nu = -1$, by using
equations~\eqref{eqt} and ~\eqref{eqfi}, we have
\begin{eqnarray}
g_{rr}\dot{r}^2 + g_{\theta\theta}\dot{\theta}^2
= V_{\rm eff}(r,\theta,E,L_z)\label{timelike} \, ,
\end{eqnarray}
where the effective potential $V_{\rm eff}$ is defined as 
\begin{eqnarray}\label{eq:potential0}
V_{\rm eff} = \frac{E^2 g_{\phi\phi} + 2 E L_z g_{t\phi} + L^2_z 
	g_{tt}}{g_{t\phi}^2 - g_{tt} g_{\phi\phi}} - 1  \, .
\end{eqnarray}
For circular equatorial orbits with $\theta=\pi/2$ and $\dot{r}=\ddot{r}=\dot{\theta}=0$, the radial component of the geodesic equations gives \cite{Pradhan:2018usf}
\begin{eqnarray} \label{omega:fi}
\Omega_\phi =\frac{\dot{\phi}}{\dot{t}} =\frac{- \partial_r g_{t\phi} 
	\pm \sqrt{\left(\partial_r g_{t\phi}\right)^2 
		- \left(\partial_r g_{tt}\right) \left(\partial_r 
		g_{\phi\phi}\right)}}{\partial_r g_{\phi\phi}} \, ,
\end{eqnarray}
where $+ (-)$  corresponds to co-rotating ( counter-rotating ) orbits respectively.
In the case of circular orbits, from equations~\eqref{eqt}, \eqref{eqfi} and \eqref{omega:fi}, the constant of motion for a test particle with constant radius ($r=r_0$) can be derived as follows
\begin{eqnarray}
E &=& - \frac{g_{tt} + g_{t\phi}\Omega_\phi}{
	\sqrt{-g_{tt} - 2g_{t\phi}\Omega_\phi - g_{\phi\phi}\Omega^2_\phi}} \, , \\
L_z &=& \frac{g_{t\phi} + g_{\phi\phi}\Omega_\phi}{
	\sqrt{-g_{tt} - 2g_{t\phi}\Omega_\phi - g_{\phi\phi}\Omega^2_\phi}} \, .
\end{eqnarray}
One of the methods to obtain the frequency of quasi-periodic oscillations is the relativistic precession model, in which one uses the geodesic perturbations  \cite{Stella:1997tc,Stella:1999sj,Schnittman:2005pi}.QPOs are generated by the particles rotating around the black hole within the accretion disk. By solving the perturbations of the geodesic equations, we obtain the oscillatory responses of a particle moving around the black hole in various directions. These responses effectively describe the particle's motion within the accretion disk.\\
Let us assume that the trajectory of the particle is subject to a slight perturbation in the  $r$ and $\theta$ directions, namely we  have $r=r_0\left(1+\delta_r \right) $
and $\theta=\pi/2+\delta_{\theta}$. Using equation~\eqref{timelike}, the equations for perturbations are given by \cite{Colistete:2002ka}
\begin{equation}
\label{eq:omegart}
\frac{d^2 \delta_r}{dt^2} + \Omega_r^2 \delta_r = 0, \quad
\frac{d^2 \delta_\theta}{dt^2} + \Omega_\theta^2 \delta_\theta = 0 \, ,
\end{equation}
where the frequencies of the oscillations are \cite{Merloni:1998by,Maselli:2017kic}
\begin{equation}\label{eq:omegas}
   \Omega^2_r = - \frac{1}{2 g_{rr} \dot{t}^2} 
\frac{\partial^2 V_{\rm eff}}{\partial r^2} \, , \quad
\Omega^2_\theta = - \frac{1}{2 g_{\theta\theta} \dot{t}^2} 
\frac{\partial^2 V_{\rm eff}}{\partial \theta^2} \,.
 \end{equation}
 \subsection{Innermost Stable Circular Orbit (ISCO)}
 One important parameter for timelike circular geodesics is the location of inner most stable circular orbits. These orbits define the boundary below which stable circular orbits do not exist. The location of these orbits can be found by requiring that $\frac{d^2 V_{eff}}{d r^2}=0$. The derivation of this orbit for Kerr metric has been given in \cite{10.1093/mnras/189.3.621,PhysRevD.5.814,Bardeen:1972fi,Jefremov:2015gza} and for the case of the Schwarzchild metric see \cite{Hagihara1931TheoryOT,Kaplan:2022fag}. The location of ISCO for the kerr-de sitter  solution is presented in \cite{Vahedi:2021ssf}. Also, the effect of cosmological constant on ISCO is studied in \cite{Howes1981-ev}.

\begin{figure}[h!]
\centering
	\includegraphics[width=0.4\linewidth]{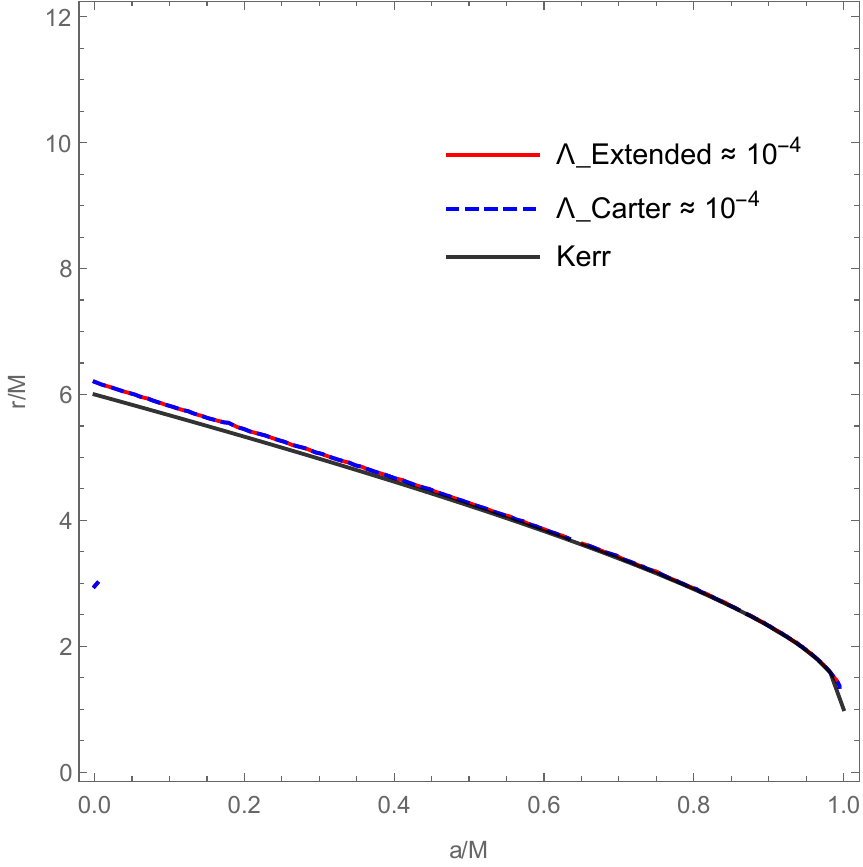}
    \includegraphics[width=0.4\linewidth]{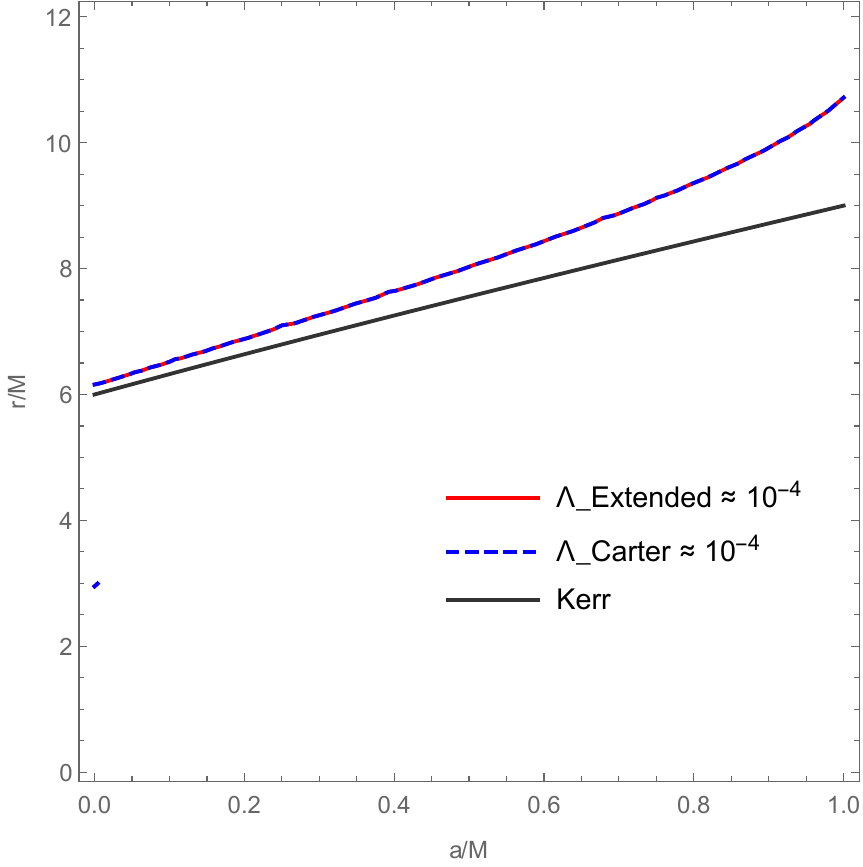} 	
	\caption{ISCO as a function of $a/M$ for prograde (left) and retrograde (right) orbits.}
	\label{fig:isco1}
\end{figure}

 The result for the extended Kerr-de sitter is presented in figure~\ref{fig:isco1}. The left (right) figure shows prograde (retrograde) orbits. We also have presented the result for Kerr and Kerr-de sitter solution for comparison.

 \subsection{ Relativistic precession model }\label{IV}
\begin{figure}
			\centering
		\begin{subfigure}{.32\linewidth}
			\centering
		\includegraphics[width=\linewidth]{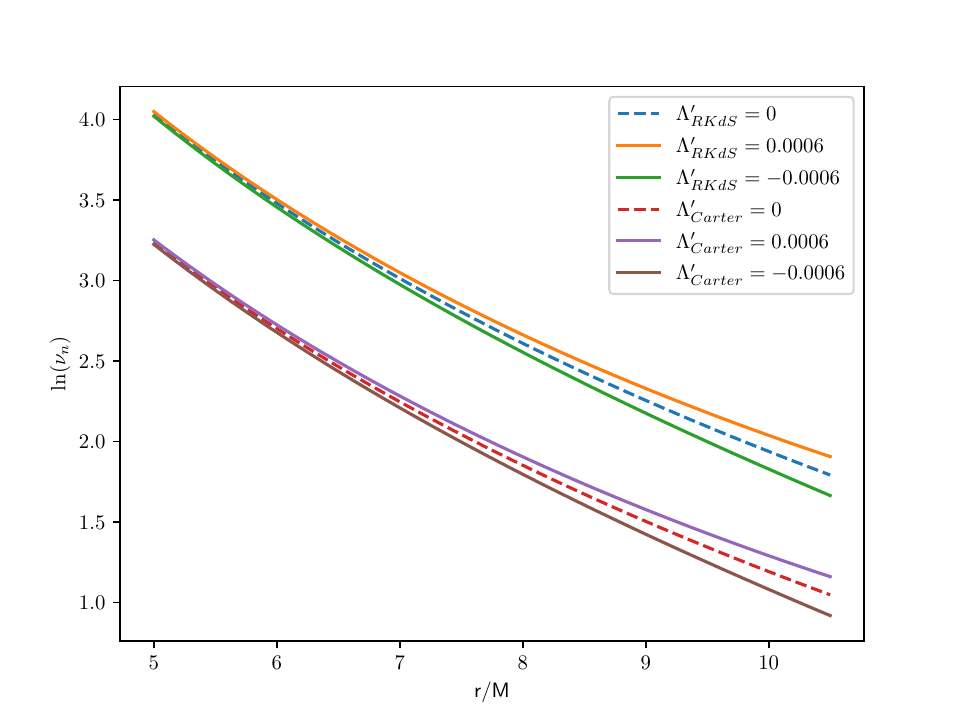}
			\caption{
				$\ln(\nu_n)$ as function of $r/M$ for various $\Lambda^\prime$ for extended Kerr-de Sitter and Carter metric.
			}
			\label{nun in terms of r GRO}
		\end{subfigure}
		\hfill
		\begin{subfigure}{.32\linewidth}
			\centering
			\includegraphics[width=\linewidth]{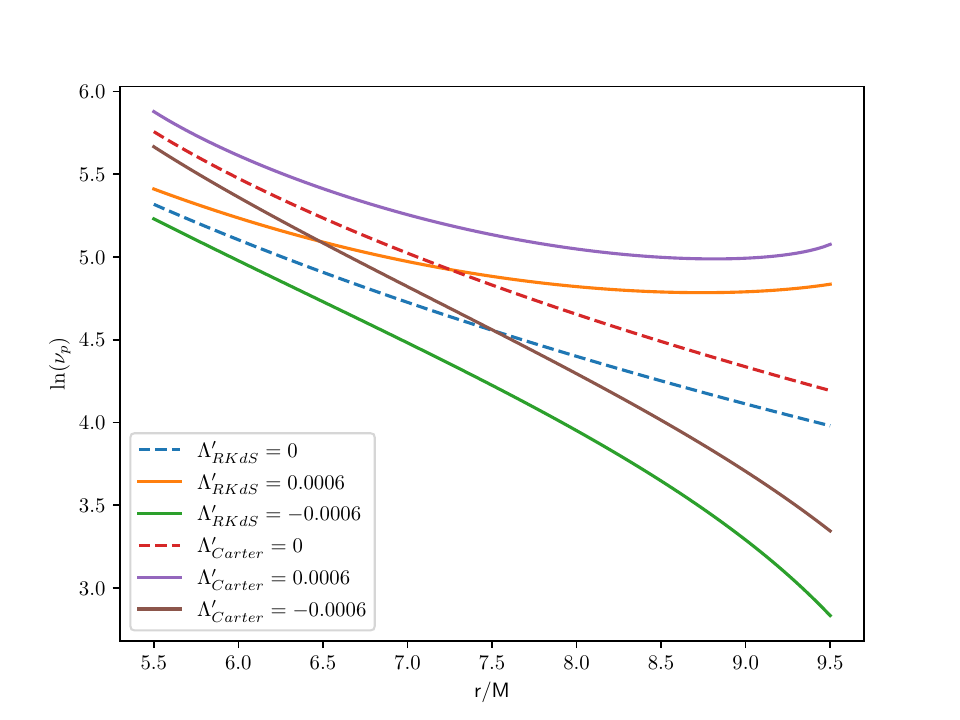}
			\caption{
				$\ln(\nu_p)$ as function of $r/M$ for various $\Lambda^\prime$ for extended Kerr-de Sitter and Carter metric.
			}
			\label{nup in terms of r GRO}
		\end{subfigure}
		\hfill
		\begin{subfigure}{.32\linewidth}
			\centering
			\includegraphics[width=\linewidth]{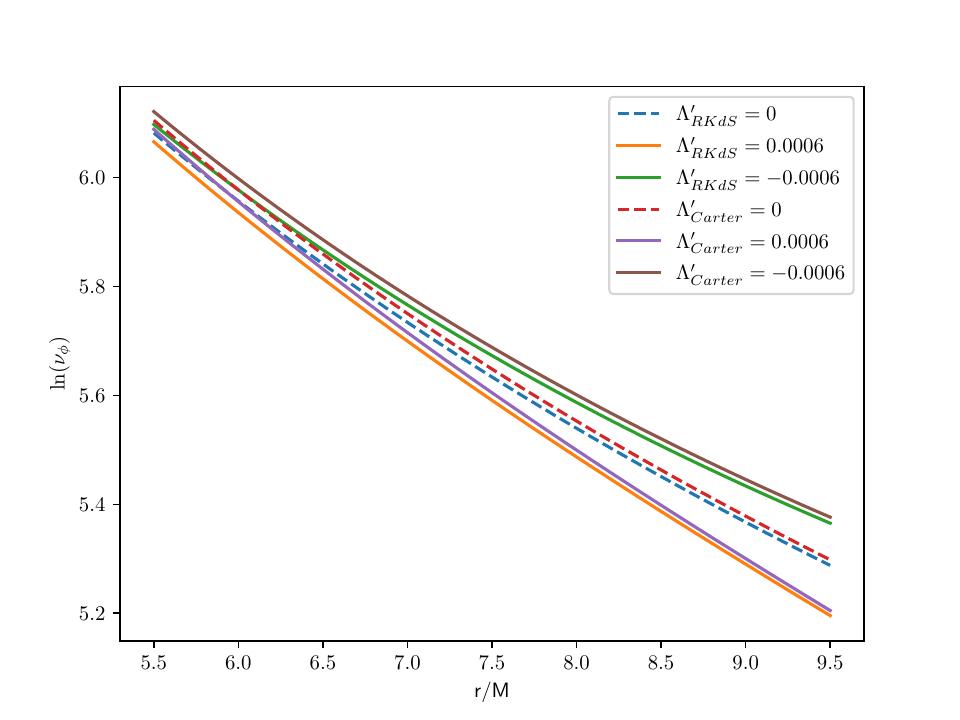}
			\caption{
				$\ln(\nu_\phi)$ as function of $r/M$ for various $\Lambda^\prime$ for extended Kerr-de Sitter and Carter metric.
			}
			\label{nuphi in terms of r GRO}
		\end{subfigure}
		\caption{$\ln(\nu_n)$,$\ln(\nu_p)$,$\ln(\nu_\phi)$ as function of $r/M$ for various $\Lambda^\prime$ for compare Carter metric and Revisited Kerr-de Sitter metric. We assumed $M/M_\odot=5.5$ and $a/M=0.3$ and used the $\Lambda^\prime$ obtained from the  Bayesian approach by the GRO J1655-40 data.}
		\label{fig:nu in terms of r GRO}
	  	\end{figure}

\begin{figure}
	\centering
	\begin{subfigure}{.32\linewidth}
	\centering
	\includegraphics[width=\linewidth]{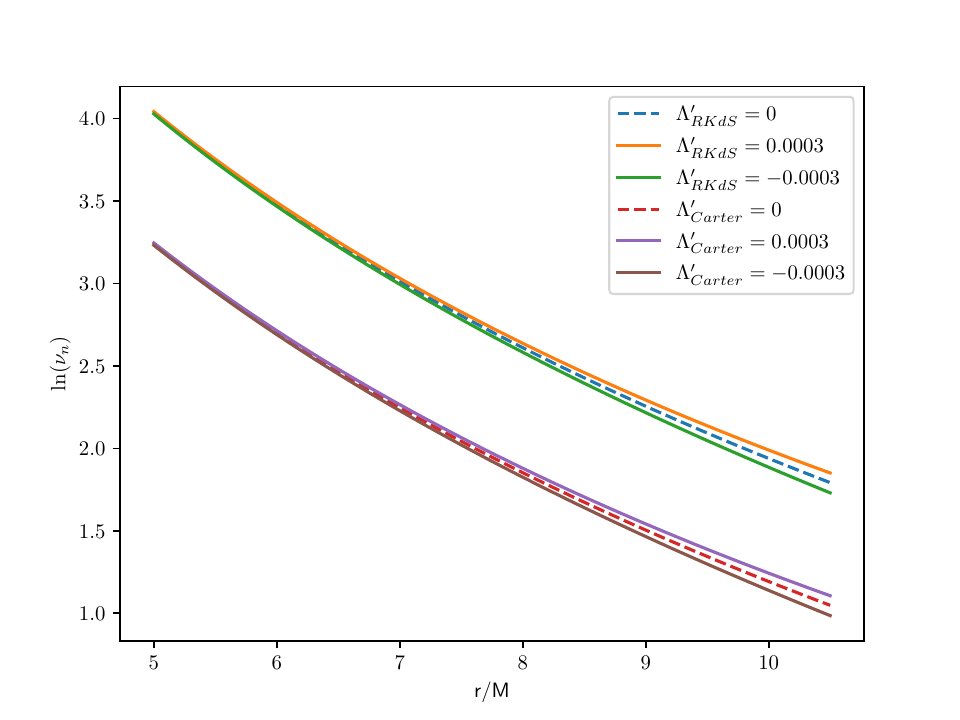}
 	\caption{$\ln(\nu_n)$ as function of $r/M$ for various $\Lambda^\prime$ for Revisited Kerr-de Sitter and Carter metric. }
 	\label{nun in terms of r XTE}	
	 \end{subfigure}
	\hfill
	\begin{subfigure}{.32\linewidth}
	\centering
	\includegraphics[width=\linewidth]{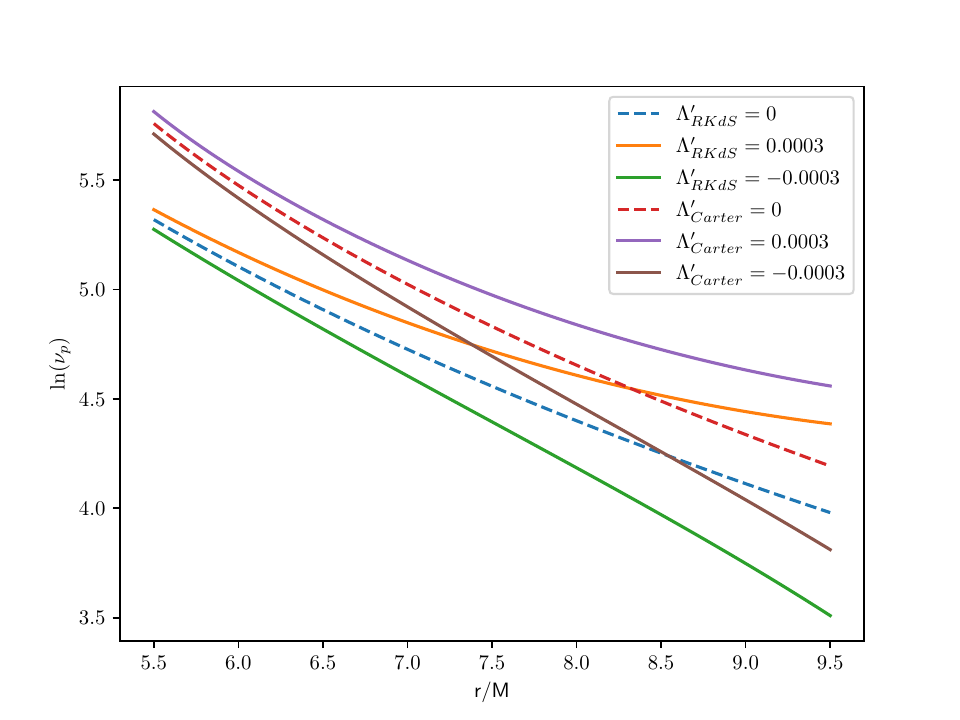}
	 \caption{$\ln(\nu_p)$ as function of $r/M$ for various $\Lambda^\prime$ for Revisited Kerr-de Sitter and Carter metric.}
	 \label{nup in terms of r XTE}
	\end{subfigure}
	\hfill
	\begin{subfigure}{.32\linewidth}
		\centering
	\includegraphics[width=\linewidth]{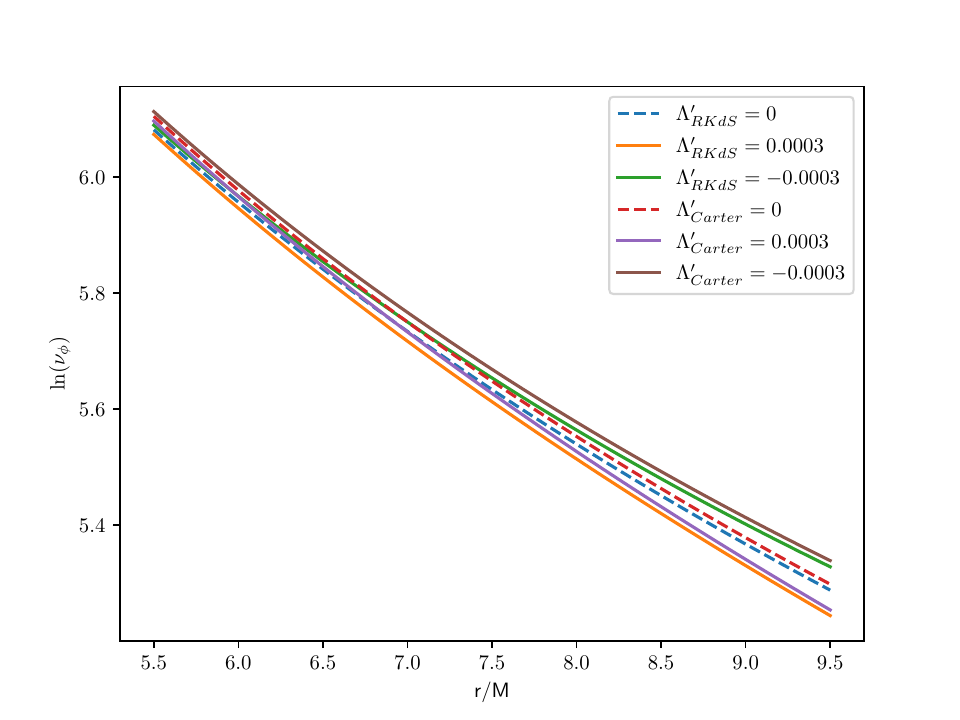} 
	\caption{$\ln(\nu_\phi)$ as function of $r/M$ for various $\Lambda^\prime$ for Revisited Kerr-de Sitter and Carter metric.}
	\label{nuphi in terms of r XTE}
	\end{subfigure}
	\caption{$\ln(\nu_n)$,$\ln(\nu_p)$,$\ln(\nu_\phi)$ as function of $r/M$ for various $\Lambda^\prime$ for compare Carter metric and Revisited Kerr-de Sitter metric. We assumed $M/M_\odot=5.5$ and $a/M=0.3$ and used the $\Lambda^\prime$ obtained from the  Bayesian approach by the XTE J1859+226 data.}
	\label{fig:nu in terms of r XTE}
     
	\end{figure}

\begin{figure}
		\centering
	\begin{subfigure}{.47\linewidth}
	\centering
	\includegraphics[width=\linewidth]{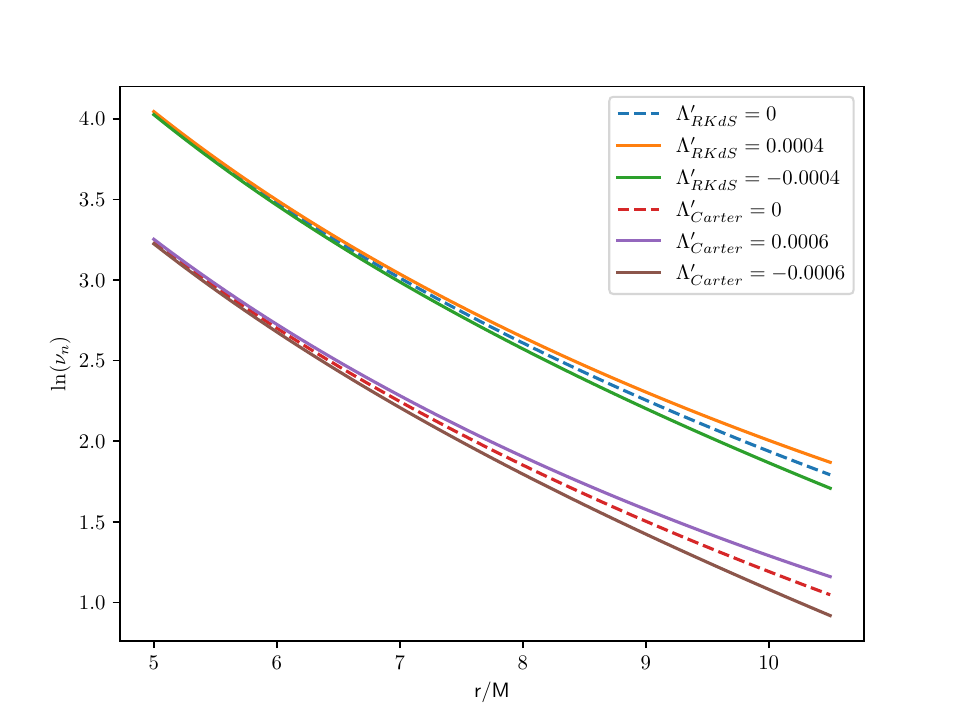}
   \caption{$\ln(\nu_n)$ as function of $r/M$ for various $\Lambda^\prime$ for Revisited Kerr-de Sitter and Carter metric. }{\label{nun in terms of r H17}}	
	\label{nun in terms of  H17}
	\end{subfigure}
	\begin{subfigure}{.47\linewidth}
		\centering
		\includegraphics[width=\linewidth]{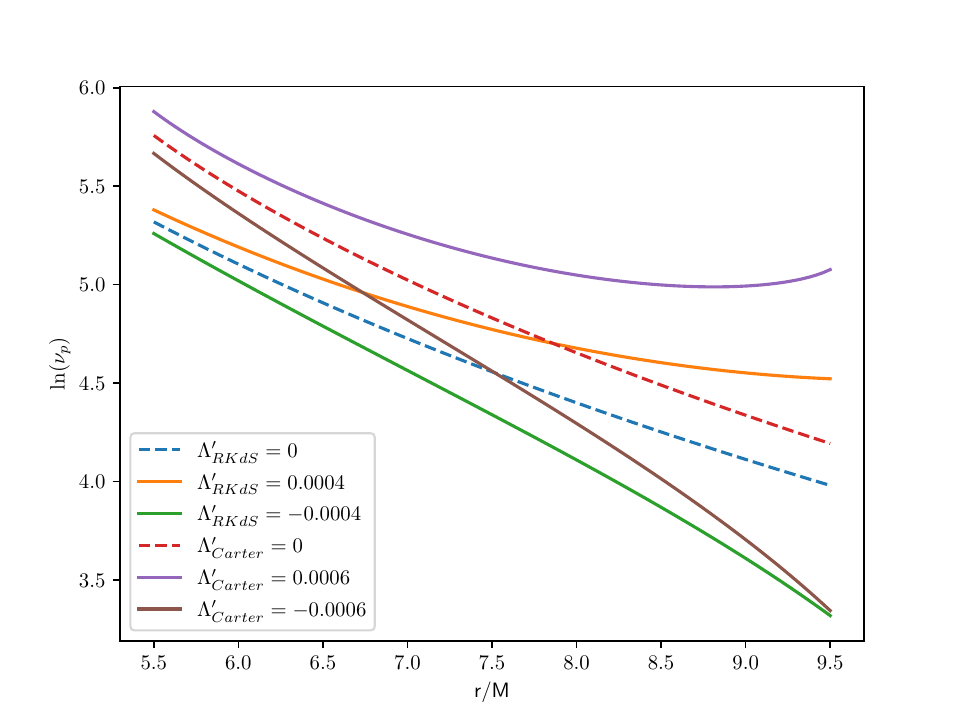}
 	\caption{$\ln(\nu_p)$ as function of $r/M$ for various $\Lambda^\prime$ for Revisited Kerr-de Sitter and Carter metric.}
	\label{nup in terms of  H17}
	\end{subfigure}
	\begin{subfigure}{.47\linewidth}
	\centering
	\includegraphics[width=\linewidth]{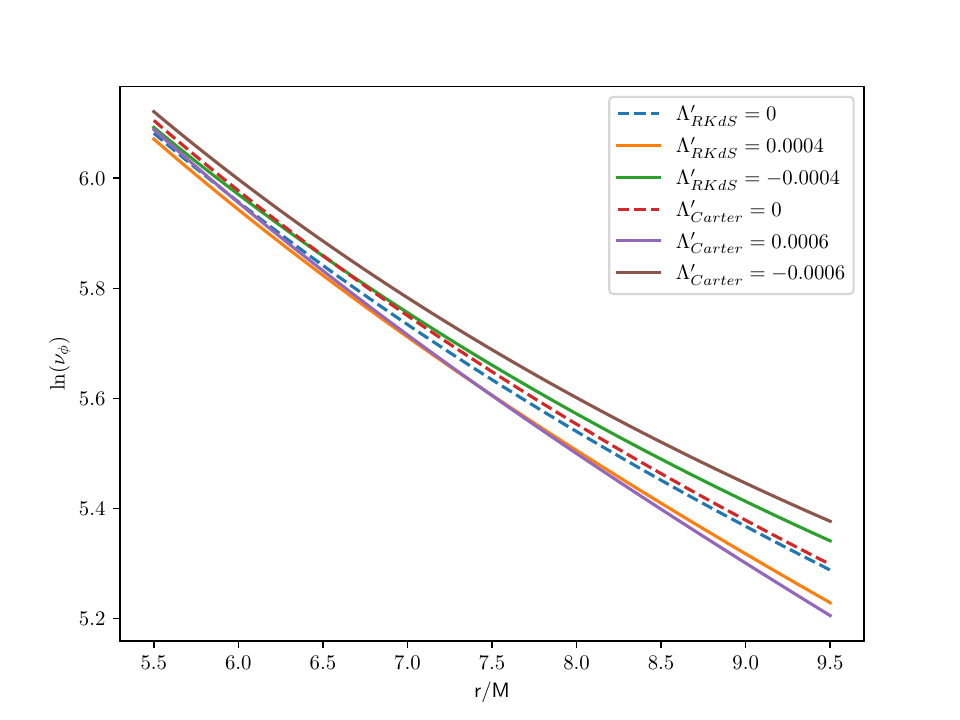}
	\caption{$\ln(\nu_\phi)$ as function of $r/M$ for various $\Lambda^\prime$ for Revisited Kerr-de Sitter and Carter metric.}
	\label{nuphi in terms of  H17}
	\end{subfigure}
	\caption{$\ln(\nu_n)$,$\ln(\nu_p)$,$\ln(\nu_\phi)$ as a function of $r/M$ for various $\Lambda^\prime$ for compare Carter metric and Revisited Kerr-de Sitter metric. We assumed $M/M_\odot=5.5$ and $a/M=0.3$ and used the $\Lambda^\prime$ obtained from the  Bayesian approach by the  H1743-322 data.}
	\label{fig:nu in terms of r H17}
	\end{figure}

As suggested by the precession model, QPOs can be correlated with the frequencies of periastron precession and nodal precession. The periastron precession
frequency  $\nu_p$ and nodal precession frequency $\nu_n$ are defined as
	\begin{equation}
		\nu_p=\nu_\phi-\nu_r , \; \nu_n=\nu_\phi-\nu_\theta,
	\end{equation}
	where $\nu_i=\Omega_i/2\pi$ and $i=n,\phi,\nu,\theta$. To show the apparent difference between the extended Kerr-de Sitter metric and the Carter Kds metric, the plots of frequencies for each metric are displayed in figures \ref{fig:nu in terms of r GRO}, \ref{fig:nu in terms of r XTE} and \ref{fig:nu in terms of r H17}. In these plots, we set $M/M_\odot=5.5$ and $a/M=0.3$. To compare with the Kerr metric, we show three cases of the dimensionless cosmological constant  $(\Lambda^\prime=\frac{\Lambda}{M^2})$ with $\Lambda^\prime=0$ and other values of $\Lambda^\prime$ are obtained from Bayesian analysis. In our study, we have used three data sets GRO J1655-40, XTE J1859+226, and H1743-322 and the results are shown in figures \ref{fig:nu in terms of r GRO}, \ref{fig:nu in terms of r XTE} and \ref{fig:nu in terms of r H17} respectively. The positive $\Lambda^\prime$ shows de Sitter space and the negative shows Anti de Sitter space. The optimal fit value of $\Lambda^\prime$ is presented in Section \ref{qpo} and summarized in Table \ref{table:result}. The figures illustrate the expected fall-off behavior of the frequencies. 
 \\
 In the next step, we compare the Kerr, Kerr-de Sitter, and extended Kerr-de Sitter metrics with experimental data.
	
\section{Constraints by QPOs }\label{qpo}

	For our purpose, we use the data available for GRO J1655$-$40 \cite{Orosz:1996cg},  XTE J1859+226 \cite{Motta:2022rku}, H1743-322 \cite{Ingram:2014ara}. These data have been observed by RXTE (Rossi X-ray Timing Explorer that was a NASA satellite that observed the time variation of astronomical X-ray sources) and 
	the method used to measure QPO frequencies is the X-ray timing method \cite{Remillard:1998ah,Strohmayer}. The data obtained from observations are depicted in Table \ref{table: data}. Utilizing a Bayesian approach, we aim to determine the optimal parameters.
Let us define the $\chi$-square as \cite{Bambi:2013fea}:
	\begin{equation}
		\label{eq:chi squred}
		\chi^2(a,\Lambda/M^2,M,r)=\frac{(\nu_C -\nu_n)^2}{\sigma_C^2}+\frac{(\nu_L -\nu_p)^2}{\sigma_L^2}+\frac{(\nu_U -\nu_\phi)^2}{\sigma_U^2},
	\end{equation}
where $\nu_C$, $\nu_L$ and $\nu_U$ are frequencies obtained from observation and $\sigma_i^2, \,(i\in \left\{C,L,U\right\})$ are their corresponding errors.
\begin{center}
		\begin{table}
   \begin{tabular}{l*{6}{c}r}
   	            & GRO J1655-40      &  \medspace \medspace \medspace    & XTE J1859+226      & \medspace \medspace  & H1743-322     &  \medspace \medspace \medspace   &       \\
   	\hline
   	$M(M_\odot)$  \medspace\medspace \medspace \medspace    & $5.4\pm0.3$ &  & $7.85\pm0.46$ &  & $\gtrsim9.29$ &  &   \\
   	$\nu_U(Hz)$                 & $441\pm2$&  & $227.5^{+2.1}_{-2.4}$ &  &  $240\pm3$ & &   \\
   	$\nu_L(Hz)$                 & $298\pm4$ &  & $128.6^{+1.6}_{-1.8}$ &  &  $165^{+9}_{-5}$ &  &    \\    
   	$\nu_C(Hz)$         & $17.3\pm0.1$ &  & $3.65\pm0.01$ & &  $9.44\pm0.02$ &  &    \\
   \end{tabular}
	\caption{Parameters $M$, $\nu_U$, $\nu_L$ and $\nu_C$ obtained from   various observations }
	\label{table: data}
	\end{table}
   	\end{center}
In Bayesian analysis, we are interested in posterior $\mathcal{P}$, the probability of the parameters given the data, which is proportional to the product of priors $p_i$ and likelihood $\mathcal{L}$:
 \begin{equation}
 	\mathcal{P}(a/M,M,\Lambda/M^2,r/M)\propto \mathcal{L}\:p(\Lambda/M^2)p(a/M)p(M)p(r),
 \end{equation}
The likelihood is also given by the Gaussian distribution. Thus we define
\begin{equation}
	\mathcal{L}\sim e^{-\frac{1}{2}\chi^2},
\end{equation}
where $\chi^2$ is defined in equation \eqref{eq:chi squred}. We study Kerr, (Carter) Kerr-de Sitter and extended Kerr-de Sitter models for each data. We use the python package \texttt{Dynesty} for sampling the posterior \cite{dynesty} and the posterior means of each case are collected in Table \ref{table:result}.
\subsection{ GRO J1655-40 data}
First, we begin with GRO J1655-40 data. The priors of the Kerr model that we set in this case, are as follows\cite{Beer:2001cg}:
\begin{equation}
0\le \frac{a}{M}\le 1,\quad 3.02\le \frac{r}{M}\le 7.02 , \quad p(M/M_\odot)\sim e^{-\frac{1}{2}\left(\frac{M/M_\odot-5.5}{0.3}\right)^2}
\end{equation}
With a uniform distribution for $\frac{a}{M}$ and $\frac{r}{M}$ and a Gaussian distribution for $M/M_\odot$.

\begin{figure}[ht!]
	\includegraphics[width=1\linewidth]{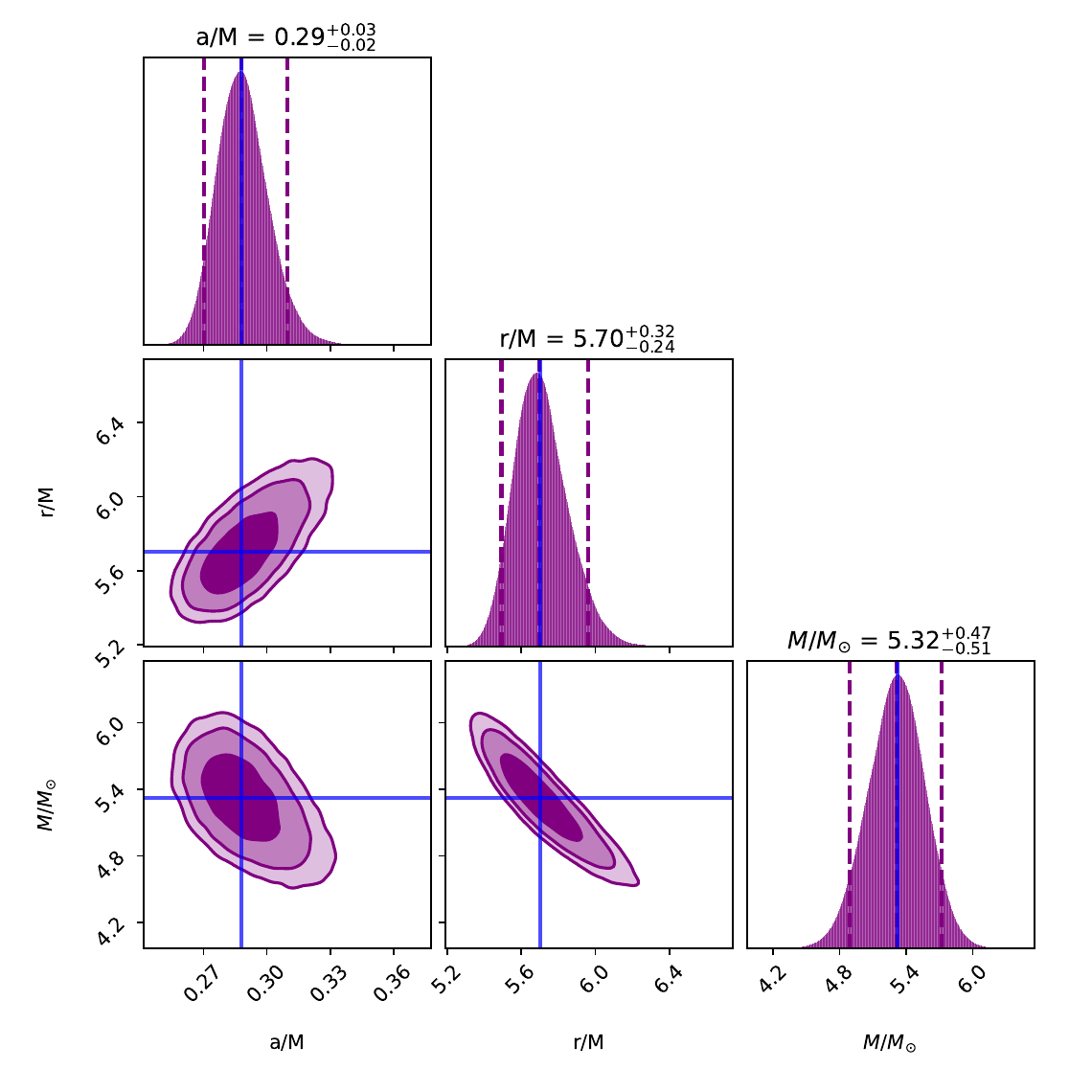}
	\centering
	\caption{The two-dimensional marginal posteriors for the parameters of the Kerr metric, derived from the GRO J1655-40 data, are displayed. Additionally, the one-dimensional distributions are represented in the diagonal plots. The contours show $ 68\% $, $ 95\% $ and
			$ 99\% $ credible intervals. Solid lines represent the mean values. Dashed lines are the $ 5\% $, $ 50\% $, and $ 95\% $
			percentiles of the distribution.}
	\label{fig:kerr daynesti GRO}
\end{figure}

The two-dimensional marginalized distribution and the covariance among the parameters are presented in Figure \ref{fig:kerr daynesti GRO} The contours show $68\%$, $ 95\% $, and $ 99\% $ credible intervals respectively.

\begin{figure}[h!]
	\includegraphics[width=.8\linewidth]{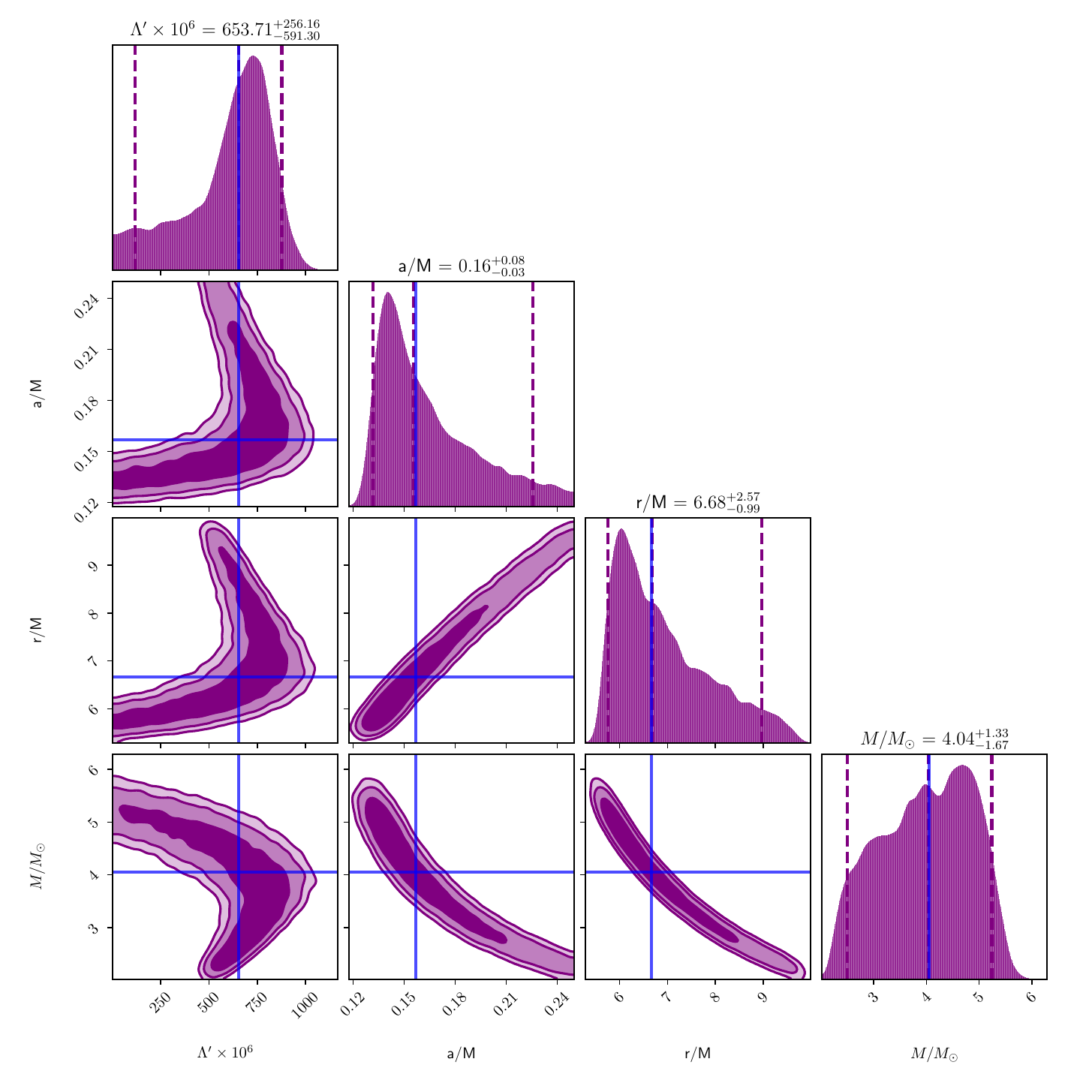}
	\centering
	\caption{The two-dimensional marginal posteriors for the parameters in the extended Kerr-de Sitter metric, based on data from GRO J1655-40, are presented here. The one-dimensional distributions are also depicted in the diagonal plots. The contours and dashed lines correspond to those shown in Figure \ref{fig:kerr daynesti GRO}.}
	\label{fig:kerrdsnew daynesti GRO}
\end{figure}

\begin{figure}[h!]
	\includegraphics[width=.8\linewidth]{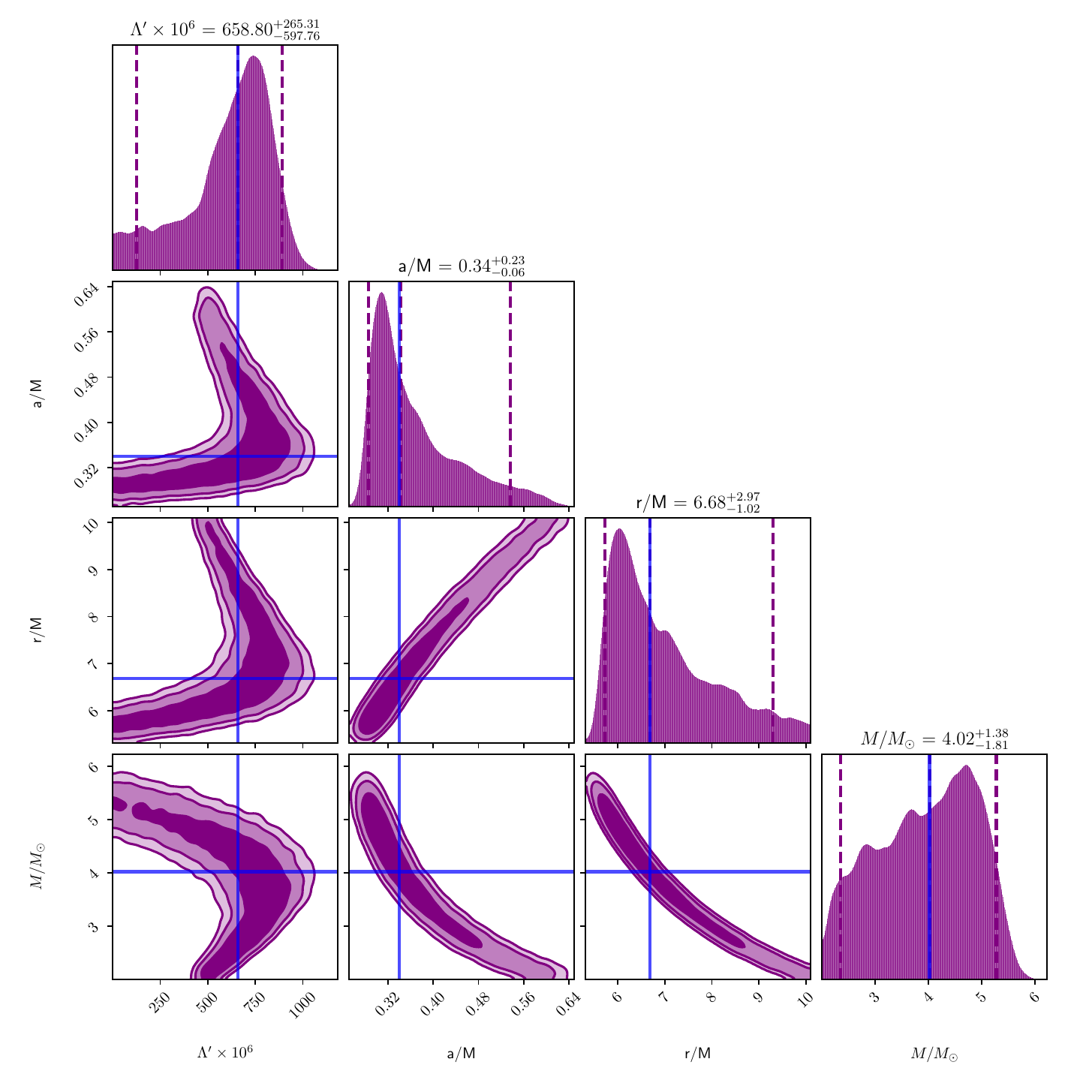}
	\centering
	\caption{The two-dimensional marginal posteriors for the parameters in the Carter-Kerr-de Sitter metric, obtained from the GRO J1655-40 data, are illustrated here. Additionally, the one-dimensional distributions are depicted in the diagonal plots. The contours and dashed lines are consistent with those presented in Figure \ref{fig:kerr daynesti GRO}.}
	\label{fig:kerrds carter daynesti GRO}
\end{figure}

For the extended Kerr-de Sitter metric, we set priors as follows
\begin{equation}
0\le \frac{\Lambda}{M^2}\le 2\times 10^{-3} ,\quad 0.05\le \frac{a}{M}\le 0.25, \quad 3.5\le \frac{r}{M}\le 11.1,\quad 2.0\le \frac{M}{M_\odot}\le 8.2.  
\end{equation}
With a uniform distribution applied to all corresponding variables.

The two-dimensional marginalized distribution and the covariance among the parameters are presented in Figure \ref{fig:kerrdsnew daynesti GRO}. The contours and dashed lines are the same as those in Figure \ref{fig:kerr daynesti GRO}. Our results indicate that it is challenging to constrain $\Lambda$ using QPO data, as $\Lambda$ is highly degenerate with other parameters. We find that our results are consistent with the Kerr description, as the precision on $\Lambda$ permits the Kerr metric even at the $ 68\% $ confidence level.

For the Carter Kerr-de Sitter metric we choose priors as follows
\begin{equation}
0\le \frac{\Lambda}{M^2}\le 2\times 10^{-3} ,\quad 0.05\le \frac{a}{M}\le 0.65, \quad 4.1\le \frac{r}{M}\le 10.1,\quad 2.0\le \frac{M}{M_\odot}\le 8.2.  
\end{equation}
With a uniform distribution for all of the corresponding variables.

The two-dimensional marginalized distribution and the covariance among the parameters are illustrated in Figure \ref{fig:kerrds carter daynesti GRO}. The contours and dashed lines are identical to those in Figure \ref{fig:kerr daynesti GRO}. Once again, our analysis reveals that constraining $\Lambda$ using QPO data proves to be challenging.

 \subsection{ XTE J1859+226}
Our next step is analyzing XTE J1859+226 data. The priors of the Kerr model that we set in this case, are as follows
\begin{equation}
	0\le \frac{a}{M}\le 1,\quad 3.02\le \frac{r}{M}\le 7.02 , \quad 2.0\le \frac{M}{M_\odot}\le 8.2.
\end{equation}
With a uniform distribution for $\frac{a}{M}$ and $\frac{r}{M}$ and a Gaussian distribution for $M/M_\odot$.
	
\begin{figure}[h!]
\includegraphics[width=1\linewidth]{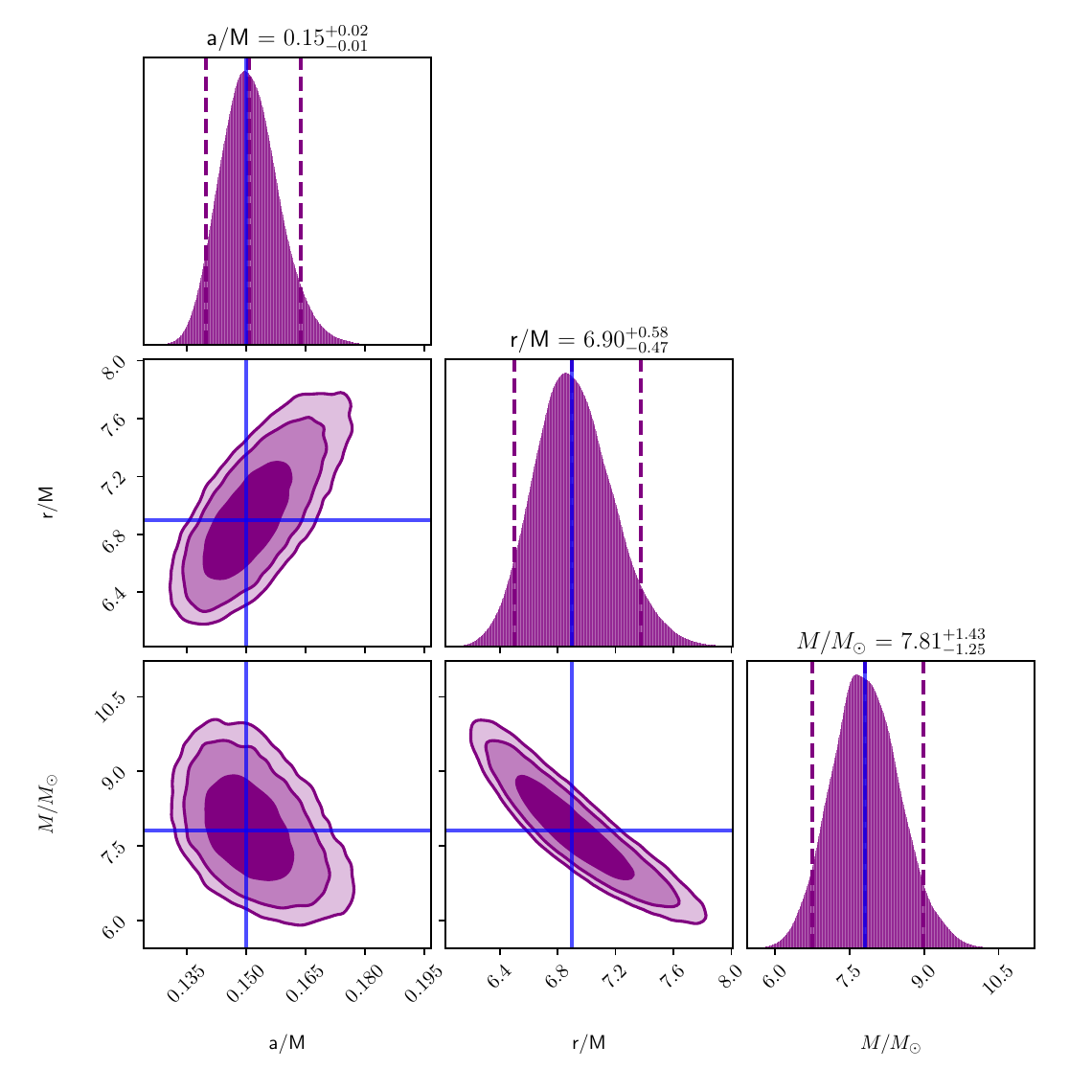}
\centering
\caption{Two-dimensional marginal posteriors for parameters in the Kerr metric by  XTE J1859+226 data. The one-dimensional distributions are also shown by the diagonal figures. The contours and dashed lines are the same as figure \ref{fig:kerr daynesti GRO}.}
\label{fig:kerr XTE}
\end{figure}

The two-dimensional marginalized distribution and the covariance among the parameters are shown in figure \ref{fig:kerr XTE}. The contours and dashed lines are the same as figure \ref{fig:kerr daynesti GRO}.

For the extended Kerr-de Sitter metric, we set priors as follows
\begin{equation}
	0\le \frac{\Lambda}{M^2}\le 2\times 10^{-3} ,\quad 0.05\le \frac{a}{M}\le 0.25, \quad 3.5\le \frac{r}{M}\le 11.1,\quad 2.0\le \frac{M}{M_\odot}\le 8.2.  
\end{equation}
With a uniform distribution for all of the corresponding variables.

\begin{figure}
	\includegraphics[width=1\linewidth]{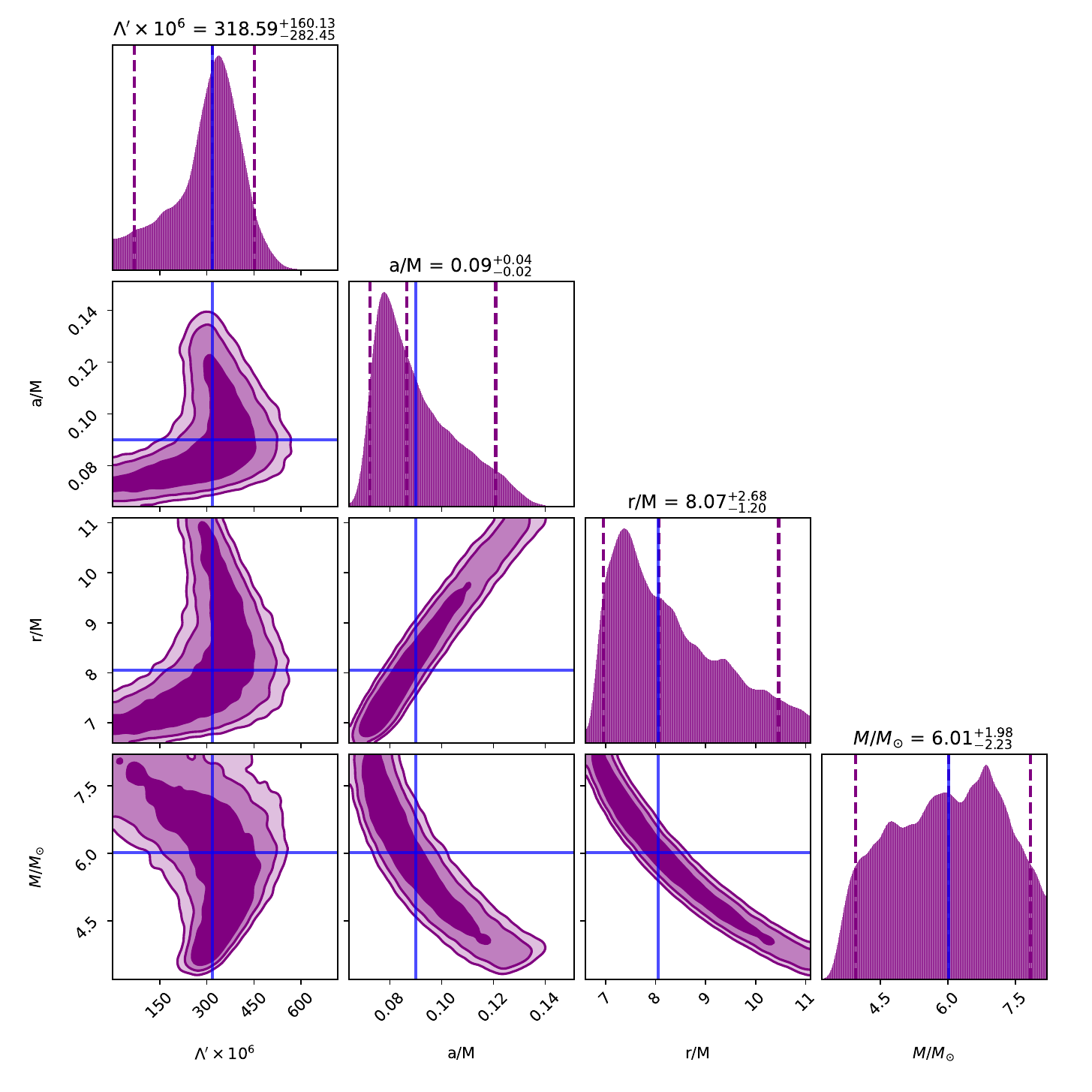}
	\centering
	\caption{The two-dimensional marginal posteriors for the parameters in the Revisited Kerr-de Sitter metric, derived from the data of XTE J1859+226, are presented here. The one-dimensional distributions are also displayed in the diagonal plots. The contours and dashed lines are consistent with those shown in Figure \ref{fig:kerr daynesti GRO}.}
	\label{fig:kerrds new XTE}
\end{figure}

The two-dimensional marginalized distribution and the covariance among the parameters are depicted in Figure \ref{fig:kerrds new XTE}. The contours and dashed lines match those presented in Figure \ref{fig:kerr daynesti GRO}. 

For Carter Kerr-de Sitter metric we choose priors as follows
\begin{equation}
	0\le \frac{\Lambda}{M^2}\le 2\times 10^{-3} ,\quad 0.05\le \frac{a}{M}\le 0.65, \quad 4.1\le \frac{r}{M}\le 10.1,\quad 2.0\le \frac{M}{M_\odot}\le 8.2.  
\end{equation}
Utilizing a uniform distribution for all relevant variables.
\begin{figure}[h!]
	\includegraphics[width=1\linewidth]{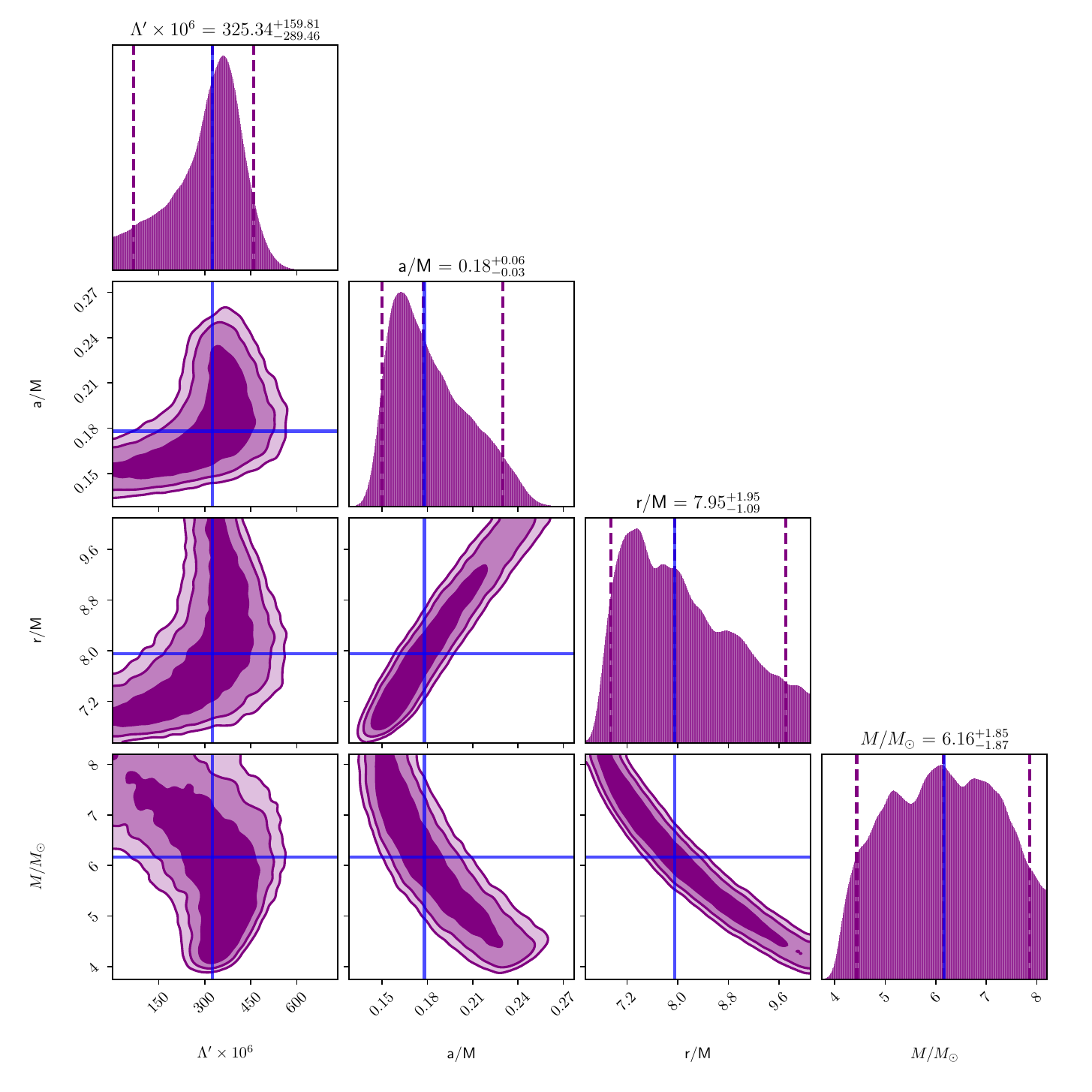}
	\centering
	\caption{Two-dimensional marginal posteriors for parameters in the Carter Kerr-de Sitter metric by XTE J1859+226 data. The one-dimensional distributions are also shown by the diagonal figures. The contours and dashed lines are the same as figure \ref{fig:kerr daynesti GRO}.}
	\label{fig:kerrds carter  XTE}
\end{figure}

The two-dimensional marginalized distribution and the covariance among the parameters are illustrated in Figure \ref{fig:kerrds carter XTE}. The contours and dashed lines are identical to those in Figure \ref{fig:kerr daynesti GRO}. 
 \subsection{ H1743-322 data}
At the final step, we analyze the data from H1743-322. The priors for the Kerr model are
\begin{equation}
	0\le \frac{a}{M}\le 1,\quad 3.02\le \frac{r}{M}\le 7.02 , \quad 9.0\le \frac{M}{M_\odot}\le 12.2.
\end{equation}
With a uniform distribution for $\frac{a}{M}$ and $\frac{r}{M}$ and a Gaussian distribution for $M/M_\odot$.

\begin{figure}[h!]
	\includegraphics[width=1\linewidth]{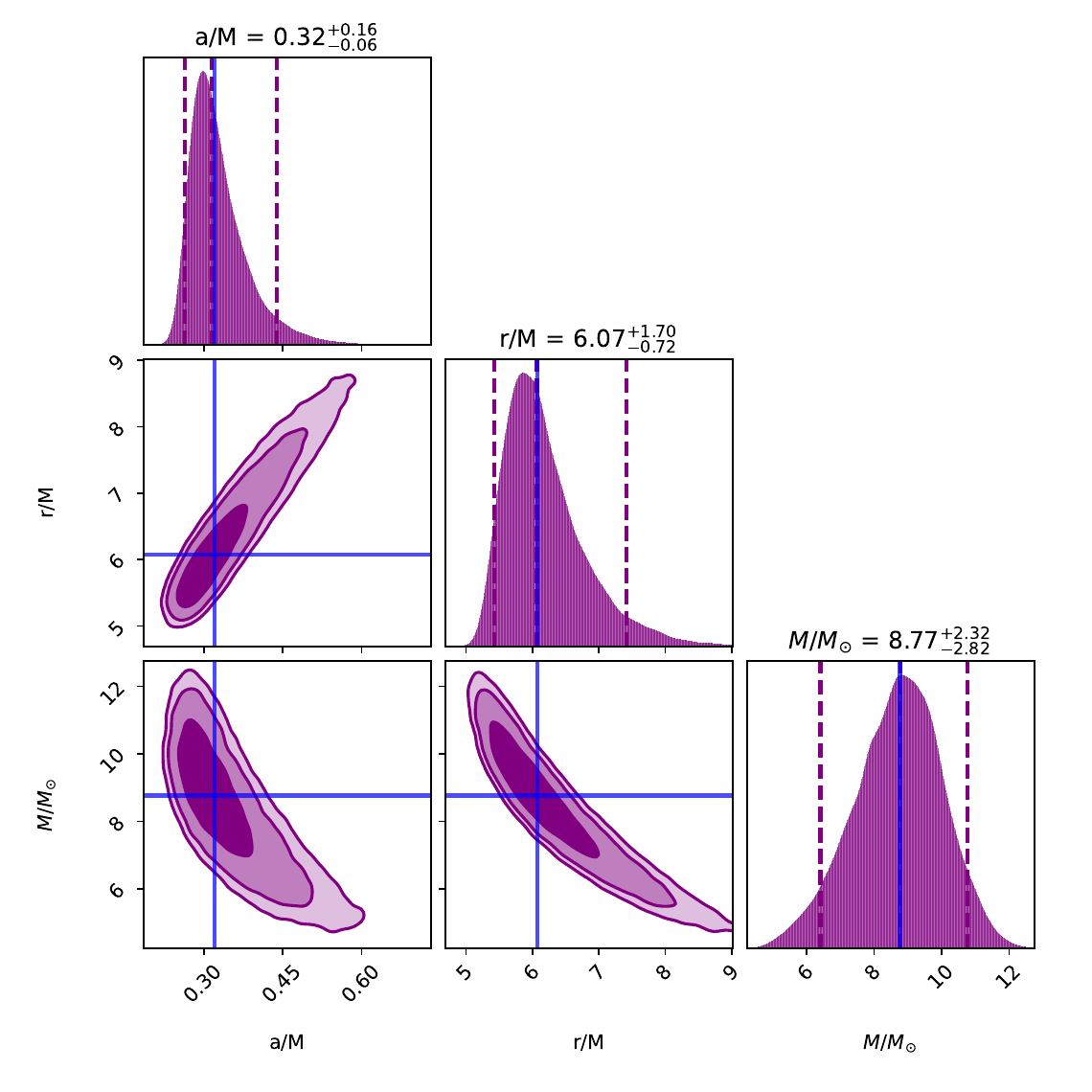}
	\centering
	\caption{Two-dimensional marginal posteriors for parameters in the Kerr metric from the data of H1743-322. The one-dimensional distributions are also shown by the diagonal figures. The contours and dashed lines are the same as figure \ref{fig:kerr daynesti GRO}.}
	\label{fig:kerr H1743}
\end{figure}

The two-dimensional marginalized distribution and the covariance among the parameters are presented in Figure \ref{fig:kerr H1743}. The contours and dashed lines are identical to those in Figure \ref{fig:kerr daynesti GRO}.

For the extended Kerr-de Sitter metric, we set priors as 
\begin{equation}
	0\le \frac{\Lambda}{M^2}\le 2\times 10^{-3} ,\quad 0.05\le \frac{a}{M}\le 0.25, \quad 3.5\le \frac{r}{M}\le 11.1,\quad 9.0\le \frac{M}{M_\odot}\le 12.2.
\end{equation}
With a uniform distribution for all of corresponding variables.

\begin{figure}[h!]
	\includegraphics[width=1\linewidth]{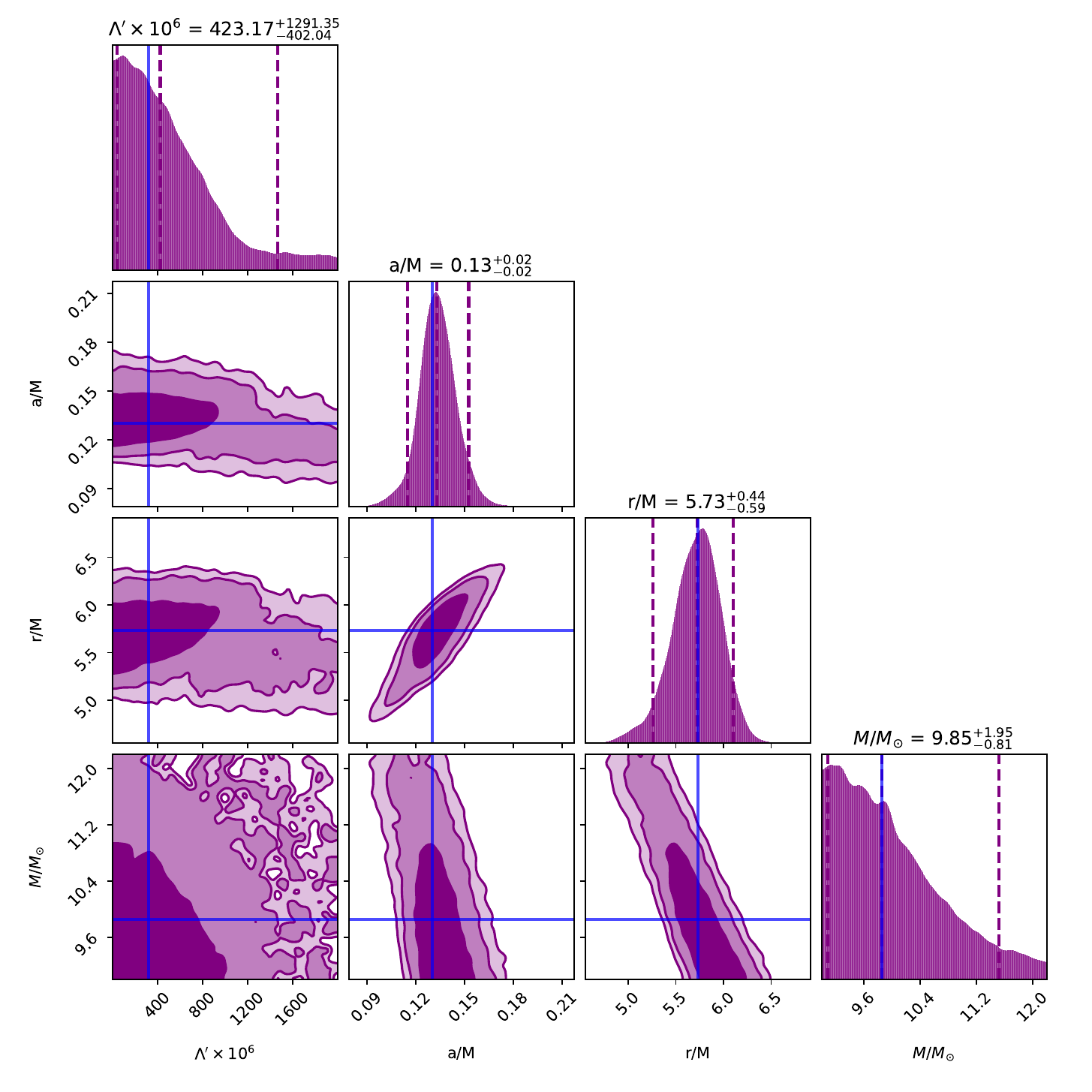}
	\centering
	\caption{Two-dimensional marginal posteriors for parameters in the extended Kerr-de Sitter metric from the data of H1743-322. The one-dimensional distributions are also shown by the diagonal figures. The contours and dashed lines are the same as figure \ref{fig:kerr daynesti GRO}.}
	\label{fig:kerrds new H1743}
\end{figure}

The two-dimensional marginalized distribution and the covariance among the parameters are shown in figure \ref{fig:kerrds new H1743}. The contours and dashed lines are the same as figure \ref{fig:kerr daynesti GRO}. 

For the Carter Kerr-de Sitter metric we choose priors as 
\begin{equation}
	0\le \frac{\Lambda}{M^2}\le 2\times 10^{-3} ,\quad 0.05\le \frac{a}{M}\le 0.65, \quad 4.1\le \frac{r}{M}\le 10.1,\quad 2.0\le \frac{M}{M_\odot}\le 8.2.  
\end{equation}
With a uniform distribution for all corresponding variables.

\begin{figure}[h!]
	\includegraphics[width=1\linewidth]{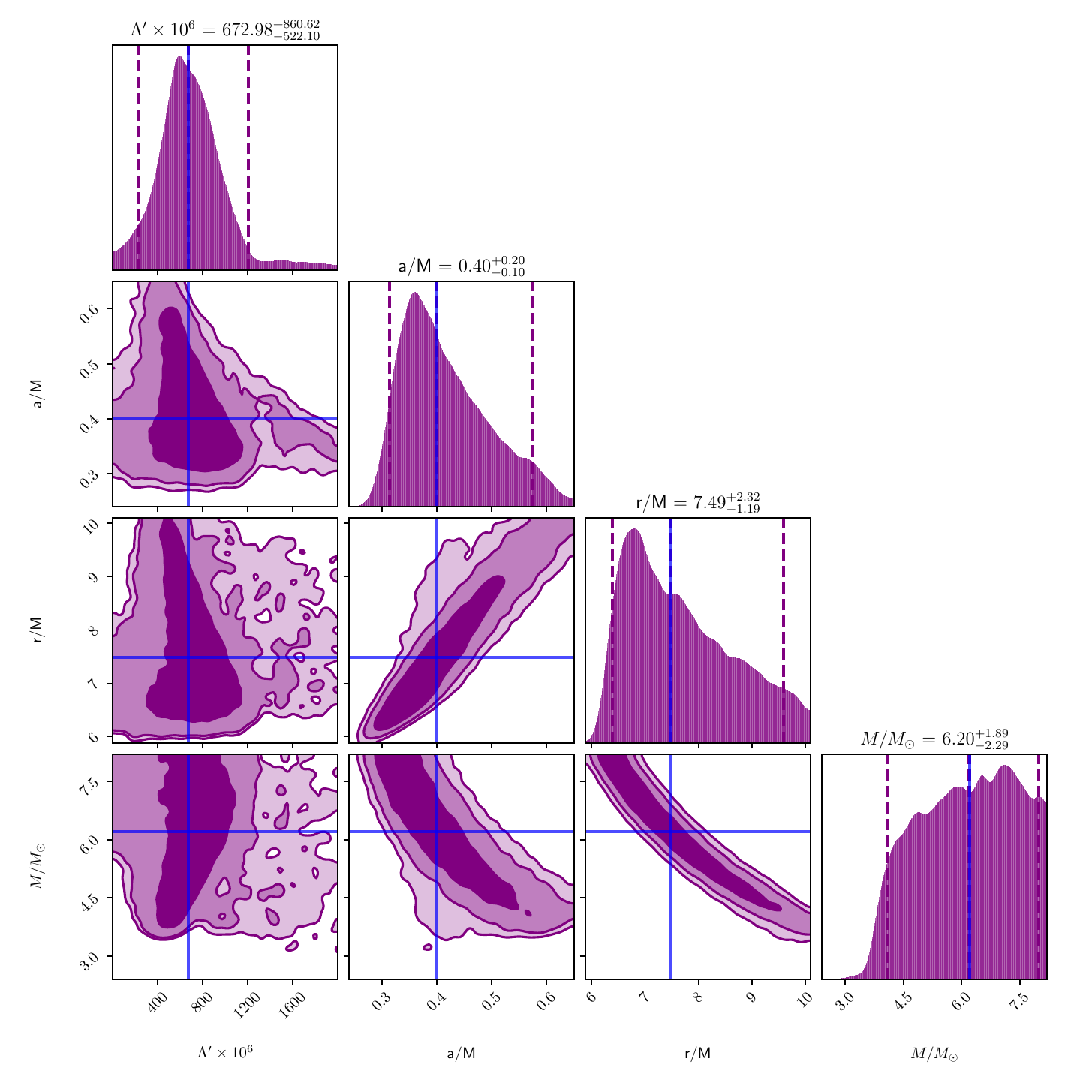}
	\centering
	\caption{Two-dimensional marginal posteriors for parameters in the Carter Kerr-de Sitter metric from the data of H1743-322. The one-dimensional distributions are also shown by the diagonal figures. The contours and dashed lines are the same as figure \ref{fig:kerr daynesti GRO}.}
	\label{fig:kerrds carter  H1743}
\end{figure}

The two-dimensional marginalized distribution and the covariance among the parameters are shown in figure \ref{fig:kerrds carter  H1743}. The contours and dashed lines are the same as figure \ref{fig:kerr daynesti GRO}. The results are summarized in table~\ref{table:result}.


\begin{center}
	\begin{table}

	\begin{tabular}{|c|c|c|c|c|c|}
		\hline
		\multicolumn{2}{|c|}{Data} 	&$a/M$ & $r/M$&$M/M_{\odot}$ & $\Lambda^{\prime}\times 10^{-6}$  \\ 
		\hline
		 \multirow{3}{*}{GRO J1655-40}&Kerr  &$0.29^{+0.03}_{-0.02}$  &$5.70^{+0.32}_{-0.24}$  &$5.32^{+0.47}_{-0.51}$ &$-$ \\  
		\cline{2-6} 
		&Kerr-de Sitter Carter &$0.34^{+0.23}_{-0.06}$ &$6.68^{+2.97}_{-1.02}$ &$4.02^{+1.38}_{-1.81}$&$658.80^{+265.31}_{-597.76}$  \\ 
		\cline{2-6} 
				&extended Kerr-de Sitter &$0.16^{+0.08}_{-0.03}$ &$6.68^{+2.57}_{-0.99}$ &$4.04^{+1.33}_{1.67}$& $653.71^{+256.16}_{-591.30}$\\

		\hline
		 \multirow{3}{*}{XTE J1859+226}& Kerr &$0.15^{+0.02}_{-0.01} $ & $6.90^{+0.58}_{-0.47}$ &$7.81^{1.43}_{-1.25}$& $-$ \\  
		 \cline{2-6} 
		 &Kerr-de Sitter Carter &$0.18^{+0.06}_{-0.03}$ & $7.95^{+1.95}_{-1.09}$&$6.16^{+1.85}_{-1.87}$&$325.34^{+159.81}_{-289.46}$ \\ 
		 \cline{2-6} 
		 &extended Kerr-de Sitter &$0.09^{+0.04}_{-0.02}$ & $8.07^{+2.68}_{-1.20}$&$6.01^{+1.98}_{-2.23}$&$318.59^{+160.13}_{-282.45}$ \\
		 \hline

		 \multirow{3}{*}{H1743-322}&Kerr  &$0.32^{+0.16}_{-0.06}$  &$6.07^{+1.70}_{-0.72}$  &$8.77^{+2.32}_{-2.82}$&$-$ \\  
		 \cline{2-6} 
		 &Kerr-de Sitter Carter &$0.40^{+0.20}_{-0.10}$ &$7.49^{+2.32}_{-1.19}$ &$6.20^{+1.89}_{-2.29}$&$672.98^{+860.62}_{-522.10}$ \\ 
		 \cline{2-6} 
		 &extended Kerr-de Sitter &$0.13^{+0.02}_{-0.02}$ &$5.73^{+0.44}_{-0.59}$ &$9.85^{+1.95}_{-0.81}$&$423.17^{+1291.35}_{-402.04}$ \\
		 \hline
		 
	\end{tabular}
	\caption{Results calculated from the Bayesian approach for Kerr, extended Kerr-de Sitter  and Kerr-de Sitter Carter metrics according to data observation of table \ref{table: data}.}
	\label{table:result}
	\end{table}
\end{center}

\section{Model comparison}\label{V}
In the  Bayesian inference approach, we have the  posterior  $\mathcal{P}\left(\mathbf{\Theta}|\mathbf{D},\mathbf{M} \right) $ for a set of parameters $\mathbf{\Theta}$ assuming a model $\mathbf{M}$ and data $\mathbf{D}$. We have
\begin{eqnarray}
\mathcal{P}\left(\mathbf{\Theta}|\mathbf{D},\mathbf{M} \right)=\frac{P(\mathbf{D}|\mathbf{\Theta},\mathbf{M})\mathbf{\pi}(\mathbf{\Theta}|\mathbf{M})}{P(\mathbf{D}|\mathbf{M})},
\end{eqnarray}
where $\mathbf{\pi}(\mathbf{\Theta})$ is the prior on $\mathbf{\Theta}$ and $P(\mathbf{D}|\mathbf{M})$ is the probability of data given the model. We have
\begin{eqnarray}
P(\mathbf{D}|\mathbf{M})=\int P(\mathbf{D}|\mathbf{\Theta},\mathbf{M})\mathbf{\pi}(\mathbf{\Theta})d^{}\mathbf{\Theta}.
\end{eqnarray}
Similarly, we can talk about the probability of the model given the data. That is $\mathcal{P}(\mathbf{M}|D)\propto P(\mathbf{D}|\mathbf{M})\mathbf{\pi}(\mathbf{M})\propto \mathcal{Z}_{\mathbf{M}}\mathbf{\pi}_{\mathbf{M}} $ where
$\mathcal{Z}_{\mathbf{M}}=P(\mathbf{D}|\mathbf{M})$ and $\mathbf{\pi}_{\mathbf{M}}=\mathbf{\pi}(\mathbf{M})$. $\mathcal{Z}_{\mathbf{M}}$ is the model evidence and $\mathbf{\pi}_{\mathbf{M}}$ is the model prior.  To compare the  models, we can define the  Bayes factor as \cite{BayesFactors}
\begin{eqnarray}
\mathcal{R}=\frac{\mathcal{Z}_{\mathbf{M_1}}\mathbf{\pi}_{\mathbf{M_1}}}{\mathcal{Z}_{\mathbf{M_2}}\mathbf{\pi}_{\mathbf{M_2}}}.  
\end{eqnarray}
If $\mathcal{R}>1$, there is evidence that $\mathbf{M_1}$ is better than $\mathbf{M_2}$ in describing data.
Here we select the Kerr metric for our comparison and consider flat priors for all the models. $\mathcal{Z}$ for each model is reported in table \ref{bayfactor}.
	\setlength{\tabcolsep}{14pt}
	\begin{center}
		\begin{table}
		
		\begin{tabular}{|c|c|c|c|}
			\hline
 			\multirow{2}{*}{Models}		& \multicolumn{3}{|c|}{$\mathcal{Z}$}   \\ \cline{2-4} 
			   & GRO J1655-40 &  XTE J1859+226  &  H1743-322\\ 
			\hline 
			Kerr & $4.10 \times 10^{-6}$ &  $5.63\times 10^{-6}$& $7.83 \times 10^{-6}$ \\
			\hline 
			Kerr-de Sitter carter & $1.82 \times 10^{-6}$ & $1.16 \times 10^{-6}$ &$9.18 \times 10^{-6}$\\
			\hline  
			Extended Kerr-de Sitter  & $2.27 \times 10^{-6}$  & $1.00 \times 10^{-6}$ &$4.38 \times 10^{-6}$\\
			\hline    
		\end{tabular}
	\caption{$\mathcal{Z}$ for Kerr, Kerr-de Sitter Carter and extended Kerr-de Sitter metrics according to data in various observation.}
 \label{bayfactor}
	\end{table}
	\end{center}
Our analysis suggests there are hints  that the latter two models may also offer an interpretation of the observed data. It is imperative to exercise caution in interpreting our findings, as we employed a simplified model for QPOs. In practice, complexities exist like environmental factors such as plasma properties and the optical thickness of the accreting material. These factors have the potential to influence the outcomes of the model evaluation.

\section{ Summary and discussion}
In this study, we have conducted observational tests on a newly proposed rotating black hole solution that incorporates a cosmological constant. This metric asymptotically approaches a de Sitter solution. We investigated the quasi-periodic oscillations (QPOs) associated with this metric by applying a relativistic precision model, which links QPOs to perturbations in the circular orbits of the surrounding accretion disc.

To constrain our model, we utilized three datasets for the QPOs and established parameter bounds through a Bayesian approach. However, we encountered challenges in measuring $\Lambda$ with precision, primarily due to degeneracies and correlations among the parameters. Notably, among the models considered, the Kerr metric continues to provide the most accurate description of the black hole dynamics.

In reality, various astrophysical phenomena complicate the modeling of accreting material around compact objects, thereby influencing the observed QPOs. Therefore, further theoretical and numerical frameworks are essential to elucidate the effects of environmental factors, such as thermal processes and the optical thickness of the accreting material, on the characteristics of observed QPOs.

    \bibliographystyle{JHEP}
	\bibliography{ref}

\end{document}